\documentclass[twocolumn,twoside]{IEEEtran}                 %% twocolumn (regular paper)

\usepackage{subfigure}
\usepackage{bm}
\usepackage{graphicx}
\usepackage{amsmath, amsthm, amsfonts, amssymb, amsbsy}
\usepackage{setspace}
\usepackage{graphicx}
\usepackage{pstricks,arydshln}
\usepackage{multirow}
\usepackage{booktabs}
\usepackage{stfloats}
\usepackage{caption}
\usepackage{cite}
\usepackage{epstopdf}
\usepackage{dsfont}
\usepackage{color}
\usepackage{graphicx}
\usepackage{epstopdf}
\usepackage{amsmath, amsthm, amsfonts, amssymb, amsbsy}
\usepackage{setspace}
\usepackage{pstricks,arydshln}
\usepackage{multirow}
\usepackage{booktabs}
\usepackage{stfloats}
\usepackage{subfigure}
\usepackage{caption}
\usepackage{cite}
\usepackage{xcolor}
\usepackage{epstopdf}
\usepackage{cases}
\usepackage[noend]{algpseudocode}
\usepackage{algorithmicx,algorithm}
\usepackage{flushend,cuted}

\newcommand{\RNum}[1]{\uppercase\expandafter{\romannumeral #1\relax}}
\begin{document}

\title{\huge AoA Estimation for OAM Communication Systems With Mode-Frequency Multi-Time ESPRIT Method}%

\author{Wen-Xuan Long,~\IEEEmembership{Graduate Student Member,~IEEE,}~Rui Chen,~\IEEEmembership{Member,~IEEE,}\\Marco Moretti,~\IEEEmembership{Member,~IEEE} and Jiandong Li,~\IEEEmembership{Fellow,~IEEE}
%\thanks{This work was supported in part by the Fundamental Research Funds for the Central Universities and the Innovation Fund of Xidian University.}
\thanks{Copyright (c) 2015 IEEE. Personal use of this material is permitted. However, permission to use this material for any other purposes must be obtained from the IEEE by sending a request to pubs-permissions@ieee.org.}
\thanks{This work was supported in part by Natural Science Basic Research Program of Shaanxi (Program No. 2021JZ-18), Natural Science Foundation of Guangdong Province of China under Grant 2021A1515010812, the open research fund of National Mobile Communications Research Laboratory, Southeast University under Grant number 2021D04, the Fundamental Research Funds for the Central Universities, and the Innovation Fund of Xidian University.}
\thanks{Wen-Xuan Long is with the State Key Laboratory of Integrated Service Networks (ISN), Xidian University, Shaanxi 710071, China, and also with the University of Pisa, Dipartimento di Ingegneria dell'Informazione, Italy (e-mail: wxlong@stu.xidian.edu.cn).}
\thanks{Rui Chen is with the State Key Laboratory of Integrated Service Networks (ISN), Xidian University, Shaanxi 710071, China, and also with the National Mobile Communication Research Laboratory, Southeast University, Nanjing 210018, China (e-mail: rchen@xidian.edu.cn).}
\thanks{Marco Moretti is with the University of Pisa, Dipartimento di Ingegneria dell'Informazione, Italy (e-mail: marco.moretti@iet.unipi.it).}
\thanks{Jiandong Li is with the State Key Laboratory of Integrated Service Networks (ISN), Xidian University, Shaanxi 710071, China (e-mail: jdli@mail.xidian.edu.cn).}
}

\maketitle

\thispagestyle{empty}
% ==================================================
\begin{abstract}
Radio orbital angular momentum (OAM) communications require accurate alignment between the transmit and receive beam directions. Accordingly, a key feature of OAM receivers is the ability to reliably estimate the angle of arrival (AoA) of multi-mode OAM beams. Considering the limitations of existing AoA estimation techniques, in this paper, we propose an easier-to-implement AoA estimation method based on applying multiple times the estimating signal parameters via rotational invariance techniques (ESPRIT) algorithm to the received training signals in OAM mode and frequency domains, which is denoted as the mode-frequency (M-F) multi-time (MT)-ESPRIT algorithm. With this method, the misalignment error of real OAM channels can be greatly reduced and the performance approaches that of ideally aligned OAM channels.
\end{abstract}


\vspace{0.0cm}
\section{Introduction}
\vspace{0.2cm}

Since the discovery in 1992 that vortex light beams can carry orbital angular momentum (OAM) \cite{Allen1992Orbital}, a significant research effort has been focused on vortex electromagnetic (EM) waves \cite{Chen2020Orbital}. The phase front of a wave carrying OAM rotates with azimuth exhibiting a helical structure $e^{j\ell\phi}$ in space, where $\phi$ is the transverse azimuth and $\ell$ is an unbounded integer defined as OAM \emph{topological charge} or OAM \emph{mode number} \cite{Allen1992Orbital}. Due to inherent orthogonality among different OAM modes, OAM wireless communications represent a novel approach for multiplexing a set of orthogonal signals on the same frequency channel to achieve the high spectral efficiency \cite{Yan2014High,Chen2018A,Cheng2018,Chen2019On,Chen2020Multi}. However, there are still some technical challenges for the practical application of OAM wireless communications.

\begin{figure}[tb] %figure1
\setlength{\abovecaptionskip}{0.2cm}   %调整图片标题与图距离
\setlength{\belowcaptionskip}{-0.0cm}   %调整图片标题与下文距离
\footnotesize
\begin{center}
\includegraphics[width=8.6cm,height=8.3cm]{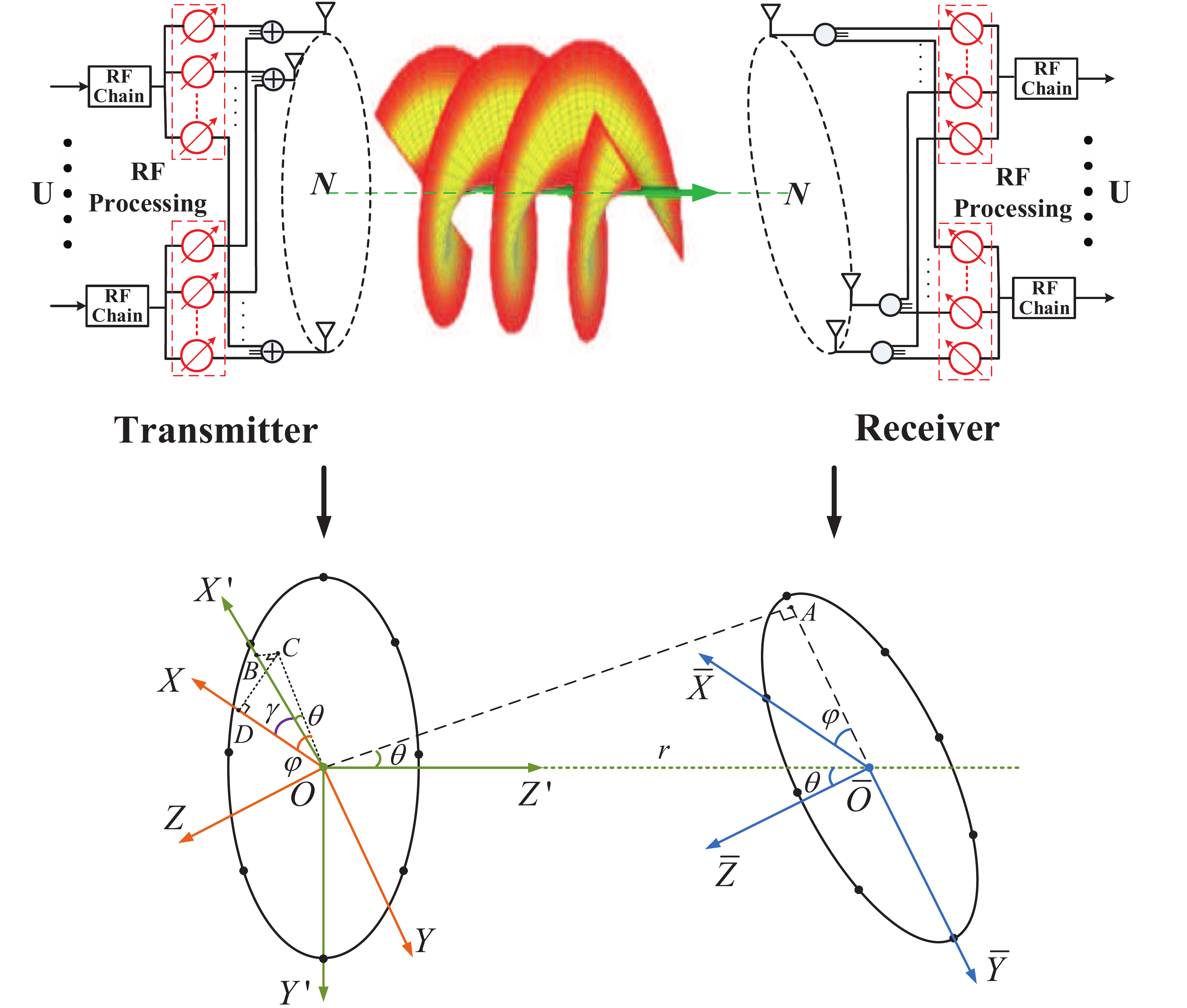}
\end{center}
\caption{The OAM communication system model and the geometrical relationship between the transmit and receive UCAs in the non-parallel misalignment case.}
\label{Fig2}
\end{figure}

One special challenge for OAM wireless communications is that they require accurate alignment between the transmit and receive antennas or at least accurate alignment between the transmit and receive beam directions \cite{Chen2018Beam}. If this condition is not precisely met, the system performance quickly deteriorates, so that the accurate estimation of the angle of arrival (AoA) is essential for OAM wireless communication systems. Although AoA estimation of planar waves has been well studied \cite{Wang2017Nuclear,Wen2017Angle,Wan2018Joint}, AoA estimation for OAM beams is still an open field of investigation. To the best of our knowledge, \cite{Chen2020Multi} is the only work addressing the problem of AoA estimation for OAM wireless communication systems. One major shortcoming of \cite{Chen2020Multi} is that it needs to process the amplitude of the received OAM training signals exploiting the knowledge of the parameters of transmit antennas, which may be difficult to obtain at the receiver.

\begin{figure*}[t]
\setcounter{equation}{2}
\begin{align} \label{xlk}
x_m(\ell_u,k_p)&=-j\frac{\mu_0\omega_p}{4\pi}\sum_{n=1}^{N}e^{i\ell_u\varphi_n}\int|\mathbf{d}_{mn}|^{-1}
e^{ik_p|\mathbf{d}_{mn}|}d\widetilde{V}_n s(\ell_u,k_p)+z_m(\ell_u,k_p)\nonumber\\
&=-j\frac{\mu_0\omega_p}{4\pi}\sum_{n=1}^{N}e^{i\ell_u\varphi_n}\int|\mathbf{r}-\mathbf{\widetilde{r}}_n+\mathbf{\widetilde{r}}_m|^{-1}
e^{ik_p|\mathbf{r}-\mathbf{\widetilde{r}}_n+\mathbf{\widetilde{r}}_m|}d\widetilde{V}_n s(\ell_u,k_p)+z_m(\ell_u,k_p)\nonumber\\
&\approx e^{i\mathbf{k}_p\cdot\mathbf{\tilde{r}}_m}\bigg[-j\frac{\mu_0\omega_p d}{4\pi}\frac{e^{ik_pr}}{r}\sum_{n=1}^{N} e^{-i(\mathbf{k}_p\cdot\mathbf{\tilde{r}}_n-\ell_u\varphi_n)} s(\ell_u,k_p) \bigg]+z_m(\ell_u,k_p)\nonumber\\
&\approx e^{ik_pR\sin\theta\cos(\varphi-\varphi_m)}\bigg[-j\frac{\mu_0\omega_p dN i^{-\ell_u}}{4\pi} \frac{e^{ik_pr}}{r} e^{i\ell_u\gamma}{J_{\ell_u}}(k_pR\sin\theta)s(\ell_u,k_p)\bigg]+z_m(\ell_u,k_p),
\end{align}
\setcounter{equation}{3}%
\hrulefill
\setlength{\belowcaptionskip}{0.0cm}   %调整图片标题与下文距离
\end{figure*}

The major contribution of this paper is to propose an easier-to-implement AoA estimation method for OAM communication systems, which applies multiple times the estimating signal parameters via rotational invariance techniques (ESPRIT) algorithm to the received training signals in OAM mode and frequency domains, thus being denoted as the mode-frequency (M-F) multi-time (MT)-ESPRIT algorithm. Simulation results validate that the proposed method is indeed able to accurately and reliably estimate the AoA of multi-mode OAM beams.

\vspace{0.2cm}
\section{OAM Communication System Model}
\vspace{0.4cm}

Employing uniform circular arrays (UCA) is a popular way to generate and receive radio OAM beams due to simple structure and multi-mode multiplexing ability \cite{Chen2018A}. Therefore, we consider a UCA-based OAM communication system, as shown in Fig.\ref{Fig2}, where the arrays at the transmitter and receiver are equipped with $N$ elements, and the transmitted signals are multiplexed on  $U$ ($1<U\leq N$) OAM modes at $P$ subcarriers. To generate the $\ell$-th mode OAM component, the $N$ elements of the transmit UCA are fed with the same input signal but with successive phase shifts ${\phi_n}=\ell\varphi_n=2{\pi}(n-1){\ell}/N, n = 1, 2, \ldots, N$,  so that after a full turn the phase has the increment of $2\pi\ell$. For an arbitrary point $\bar{P}(\bar{r},\bar{\varphi},\bar{\theta})$ in the far field, the electric field intensity $E(\mathbf{\bar{r}},k,\ell)$ generated by the transmitter can be written as \cite{Mohammadi2010Orbital,Chen2019On}
%at $\textrm{Z}'-\textrm{X}'\textrm{O}\textrm{Y}'$ coordinate system shown in Fig.\ref{Fig2},
\setcounter{equation}{0}
\begin{align} \label{E}
E(\mathbf{\bar{r}},k,\ell)&=-j\frac{\mu_0\omega}{4\pi}\sum_{n=1}^{N}e^{i\ell\varphi_n}\int|\mathbf{\bar{r}}-\mathbf{\widetilde{r}}_n|^{-1}
e^{ik|\mathbf{\bar{r}}-\mathbf{\widetilde{r}}_n|}d\widetilde{V}_n \nonumber\\
&\overset{(a)}\approx-j\frac{\mu_0\omega d}{4\pi}\frac{e^{ik\bar{r}}}{\bar{r}}\sum_{n=1}^{N}e^{-i(\mathbf{k}\cdot\mathbf{\tilde{r}}_n-\ell\varphi_n)}\nonumber\\
&\overset{(b)}\approx-j\frac{\mu_0\omega dNe^{ik\bar{r}}e^{i\ell\bar{\varphi}}}{4\pi \bar{r}}i^{-\ell}{J_\ell}(kRsin\bar{\theta}),
\end{align}
where $j$ is the amplitude of the constant current density of the dipole, $i$ is the imaginary unit, $\mu_{0}$ is the magnetic conductivity in the vacuum, $d$ is the electric dipole length, $\omega=2\pi f$ is the circular frequency, $\mathbf{k}$ is the wave vector and $|\mathbf{k}|=k=\frac{2\pi c}{f}$, $c$ is the speed of light in vacuum and $f$ is the frequency, $J_{\ell}(\cdot)$ is the $\ell$th-order Bessel function of the first kind, $R$ is the radius of UCA. In \eqref{E}, (a) applies the approximation $|\mathbf{\bar{r}}-\mathbf{\tilde{r}}_n|\approx \bar{r}$ for amplitudes and $|\mathbf{\bar{r}}-\mathbf{\tilde{r}}_n|\approx \bar{r}-\mathbf{\hat{r}}\cdot\mathbf{\tilde{r}}_n$ for phases in the far field, where $\mathbf{\hat{r}}$ is the unit vector of $\mathbf{\bar{r}}$, $\mathbf{\tilde{r}}_n=R\left(\mathbf{x'}\cos\varphi_n+\mathbf{y'}\sin\varphi_n\right)$, $\mathbf{x'}$ and $\mathbf{y'}$ are the unit vectors of x-axis and y-axis of the coordinate system at the transmitter, respectively, and (b) holds when $N$ is large enough.

In practice, accurate alignment between the transmit and receive UCAs may be difficult to realize. For easier analysis, we consider the non-parallel misalignment case as shown in Fig.\ref{Fig2} \cite{Chen2018Beam, Chen2020Multi}, where the center of transmit OAM beams overlaps with the receive UCA center, but the receive UCA plane has an angular shift relative to the transmit UCA plane. In the geometrical model, the transmitter coordinate system $\textrm{Z}'-\textrm{X}'\textrm{O}\textrm{Y}'$ is established using the transmit UCA plane as the $\textrm{X}'\textrm{O}\textrm{Y}'$ plane and the axis through the point $\textrm{O}$ and perpendicular to the UCA plane as the $\textrm{Z}'$-axis, and the receiver coordinate system $\bar{\textrm{Z}}-\bar{\textrm{X}}\bar{\textrm{O}}\bar{\textrm{Y}}$ is established on the plane of the receive UCA by the similar approach. In the non-parallel misalignment case, the coordinates of the receive UCA center can be denoted as $\bar{\textrm{O}}(r,0,0)$ in $\textrm{Z}'-\textrm{X}'\textrm{O}\textrm{Y}'$, and the coordinates of the transmit UCA center is denoted as $\textrm{O}(r,\varphi,\theta)$ in $\bar{\textrm{Z}}-\bar{\textrm{X}}\bar{\textrm{O}}\bar{\textrm{Y}}$, where $r$ is the distance between the transmit and the receive UCA centers, $\varphi$ is the azimuth angle, $\theta$ is the elevation angle, and \emph{$\varphi$ and $\theta$ are defined as the AoA of OAM beams}.

\begin{figure*}[t]
\begin{center}
\includegraphics[scale=0.5]{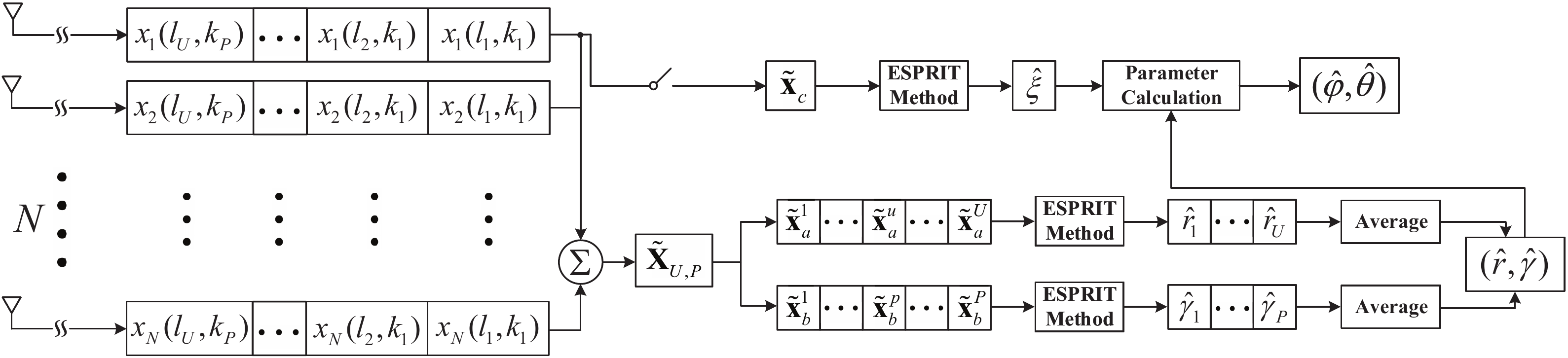}
\end{center}
\caption{The AoA estimation of the OAM wireless communication system with M-F MT-ESPRIT method}.
\label{Fig4}
\end{figure*}

To obtain the estimates of $\varphi$ and $\theta$, we have to first build the coordinate system $\textrm{Z}-\textrm{XOY}$ that is at the transmit UCA and parallel to $\bar{\textrm{Z}}-\bar{\textrm{X}}\bar{\textrm{O}}\bar{\textrm{Y}}$, and then get the expression of \eqref{E} at $\bar{\textrm{O}}$ point in $\textrm{Z}-\textrm{XOY}$ coordinate system.
According to the geometrical model, the angle between $\textrm{Z}'$-axis and $\bar{\textrm{Z}}$-axis is $\pi-\theta$, so as $\textrm{Z}$-axis and $\textrm{Z}'$-axis. Let $\textrm{B}$ be a point on the $\textrm{X}'$-axis, $\textrm{C}$ is the projection of $\textrm{B}$ on the XOY plane, and $\textrm{D}$ is the projection of $\textrm{C}$ on the $\textrm{X}$-axis, then $\angle\textrm{BOC}=\theta$. As the lines OC and $\bar{\textrm{O}}\textrm{A}$ are parallel, $\angle \textrm{COD}=\varphi$. Denote the angle between $\textrm{X}'$-axis and $\textrm{X}$-axis as $\gamma$, then $\gamma$ can be obtained as
\begin{align}\label{psi}
\gamma &= \angle \textrm{BOD} = \arccos\left(\cos(\angle\textrm{BOC})\cos(\angle \textrm{COD})\right)\nonumber\\
&= \arccos\left(\cos\theta\cos\varphi\right).
\end{align}
Based on \eqref{E} and the geometrical model, the electric field intensity at the $m$th $(1\leq m\leq N)$ element of receive UCA in the $\textrm{Z}-\textrm{X}\textrm{O}\textrm{Y}$ coordinate system can be expressed as \eqref{xlk}, where $\mathbf{d}_{mn}$ is the position vector from the $n$th transmit antenna element to the $m$th receive antenna element, $\mathbf{\tilde{r}}_m=R(\mathbf{\bar{x}}\cos\varphi_m+\mathbf{\bar{y}}\sin\varphi_m)$, $\mathbf{\bar{x}}$ and $\mathbf{\bar{y}}$ are the unit vectors of $\bar{\textrm{X}}$-axis and $\bar{\textrm{Y}}$-axis, $\varphi_m=2{\pi}(m-1)/N$, $s(\ell_u,k_p)$ is the $\ell_u$-th $(-U_L\leq \ell_u \leq U_R, U_L+U_R=U)$  OAM mode information symbol on the $u$th timeslot at frequency $f_p=k_p c/(2\pi)$ $(1\leq p \leq P)$, $z_m(\ell_u,k_p)$ is the white Gaussian noise affecting the $\ell_u$-th OAM mode at frequency $f_p$. Then, all the signals received by the $N$ elements of receive UCA in the $\textrm{Z}-\textrm{X}\textrm{O}\textrm{Y}$ coordinate system are combined, which is expressed as
\setcounter{equation}{3}
\begin{align}\label{xbc}
x(\ell_u,k_p)&=\sum_{m=1}^{N} x_m(\ell_u,k_p)\nonumber\\
&\approx\beta_{u,p}\frac{e^{ik_pr}}{r}e^{i\ell_u\gamma}{J_{\ell_u}}(k_pR\sin\theta){J_{0}}(k_pR\sin\theta)\nonumber\\
&\quad +z(\ell_u,k_p),
\end{align}
where $\beta_{u,p}=-j\frac{\mu_0\omega_p dN^2}{4\pi}i^{-\ell_u}s(\ell_u,k_p)$, $z(\ell_u,k_p)$ is the corresponding additional noise.
%$\sigma_{\ell_q,k_p}=-j\frac{\mu_0\omega_p dN^2}{4\pi r}s(\ell_q,k_p)$ is a term unrelated to $\theta$ and $\varphi$,

As it happens for multiple-input multiple-output (MIMO) communications, also OAM transmitted frames may  carry a combination of  training and data symbols. In the case of MIMO, pilot data is used to perform channel estimation, and it is well known that the number of  training symbols per subcarrier or subchannel has to be larger than the number of transmit antennas. On the other hand, in the case of OAM, pilot data is exploited for AoA estimation and only $U~(1\leq U \leq N)$ training symbols per subcarrier or subchannel are required. Accordingly, we assume that $\{s(\ell_u,k_p)|U-L\leq\ell_u\leq U_R, 1\leq p\leq P\}$ are the pilot signals in the training phase known to the OAM receiver.

\section{The AoA Estimation Based on M-F MT-ESPRIT}

\subsection{Problem Formulation}

The aim of the AoA estimation is to obtain the estimates of the azimuthal angle $\varphi$ and the elevation angle $\theta$. From \eqref{xlk} and \eqref{xbc}, we can observe that the OAM mode number $\ell_u$ and the angle $\gamma$ satisfy the dual relationship, so do $k_p$ and $r$, where $\gamma=\arccos\left(\cos\theta\cos\varphi\right)$ is an intermediate parameter related to $\varphi$ and $\theta$. Meanwhile, the elevation angle $\theta$ is associated with both $\ell_u$ and $k_p$. Therefore, we propose to first estimate $r$ and $\gamma$ with the two dual relationships in \eqref{xbc}, and then obtain the estimates of $\varphi$ and $\theta$ according to \eqref{xlk}. The working process of AoA estimation for multi-mode OAM beams is shown in Fig.\ref{Fig4}.

\subsection{Estimation of $r$ and $\gamma$}

To estimate $r$ and $\gamma$, we exploit the fact that we can obtain independent estimates of these parameters on each mode $\ell_u$ and subcarrier $k_p$. Accordingly, at the receiver, we need to extract the exponentials containing $r$ and $\gamma$ from the signal $x(\ell_u,k_p)$ by performing the following operation
\setcounter{equation}{4}
\begin{align}
\tilde{x}(\ell_u,k_p)&=-\frac{x(\ell_u,k_p)}{|x(\ell_u,k_p)|} \frac{s^*(\ell_u,k_p)}{|s(\ell_u,k_p)|}i^{\ell_u}{\rm sign}(x(\ell_u,k_p))\nonumber\\
&=e^{ik_pr}e^{i\ell_u\gamma}+\tilde{z}(\ell_u,k_p),
\label{tildex}
\end{align}
where $\tilde{z}(\ell_u,k_p)$ is the corresponding noise.

All the signals are received on the $U$ OAM modes at the $P$ frequencies and  can be collected in the matrix
\begin{equation}
\mathbf{\tilde{X}}_{U,P}=\begin{bmatrix}
\tilde{x}(\ell_1,k_1) & \tilde{x}(\ell_1,k_2) & \cdots & \tilde{x}(\ell_1,k_P) \\
\tilde{x}(\ell_2,k_1) & \tilde{x}(\ell_2,k_2) & \cdots & \tilde{x}(\ell_2,k_P) \\
    \vdots     &   \vdots        & \ddots &   \vdots   \\
\tilde{x}(\ell_U,k_1) & \tilde{x}(\ell_U,k_2) & \cdots & \tilde{x}(\ell_U,k_P) \\
\end{bmatrix}.
\label{matrixX}
\end{equation}
For easier analysis, we assume the adopted frequencies and OAM modes satisfy $k_{p+1}-k_p=1$ and $\ell_{{u}+1}-\ell_{{u}}=1$.

In the estimation of $r$, we first denote the $u$th row of $\mathbf{\tilde{X}}_{U,P}$ as a column vector $\mathbf{\tilde{x}}_{a}^u$, i.e.,
\begin{align*}
\mathbf{\tilde{x}}_{a}^u=\mathbf{\tilde{X}}_{U,P}(u,:)=[\tilde{x}(\ell_u,k_1), \tilde{x}(\ell_u,k_2),\cdots, \tilde{x}(\ell_u,k_P)]^\mathrm{T}.
\end{align*}
In the ESPRIT-based distance estimation method \cite{Chen2020Multi}, we can obtain an estimate of $r$ based on the $u$th row of the matrix $\mathbf{\tilde{X}}_{U,P}$, which we denote as $\hat{r}_u$. Hence, there are $U$ estimates of $r$ in total, which can be expressed as
\begin{align}
\hat{r}_u=&r+\varepsilon_u,  {u}=1,2,\cdots,U,\nonumber
\end{align}
where $\{\varepsilon_u|u$$=$$1,2,\cdots,U\}$ represent the estimation errors at $U$ modes. Suppose that $\{\varepsilon_u| u=1,2,\cdots,U\}$ have the same average variance $\textrm{Var}(\varepsilon_r)$, thus,
\begin{align}
\textrm{Var}\left(\frac{1}{U}\sum_{u=1}^{U}\hat{r}_u\right)= \frac{\textrm{Var}\left(\varepsilon_r\right)}{U}.
\end{align}
Therefore, $\hat{r}=\frac{1}{U}\sum_{u=1}^{U}\hat{r}_u$ is adopted as the estimate of $r$.

In the estimation of $\gamma$, we similarly denote the $p$th column of $\mathbf{\tilde{X}}_{U,P}$ as
\begin{align*}
\mathbf{\tilde{x}}_{b}^p=\mathbf{\tilde{X}}_{U,P}(:,p)=[\tilde{x}(\ell_1,k_p),\tilde{x}(\ell_2,k_p),\cdots,\tilde{x}(\ell_U,k_p)]^\mathrm{T}.
\end{align*}
After that, by following the same method employed for $\hat{r}$ we can obtain $\hat{\gamma}$, the estimate of $\gamma$.

\begin{algorithm}[t]
\label{alg1}
\caption{AoA Estimation by M-F MT-ESPRIT Algorithm}
\hspace*{0.02in} {\bf Input:}
$\mathbf{\tilde{X}}_{U,P}$ and $\mathbf{\tilde{x}}_{c}$\\
\hspace*{0.02in} {\bf Output:}
$\varphi$ and $\theta$
\begin{algorithmic}[1]
\State \textbf{procedure}
\State \quad $\mathbf{\tilde{x}}_{a}^u\leftarrow\mathbf{\tilde{X}}_{U,P}(u,:), u=1,2,\cdots,U$
\State \quad $\hat{r}_u\leftarrow$ ESPRIT algorithm for $\mathbf{\tilde{x}}_{a}^u, u=1,2,\cdots,U$
\State \quad $\hat{r} \leftarrow$ $\frac{1}{U}\sum_{u=1}^{U}\hat{r}_u$
\State \quad $\mathbf{\tilde{x}}_{b}^p\leftarrow\mathbf{\tilde{X}}_{U,P}(:,p), p=1,2,\cdots,P$
\State \quad $\hat{\gamma}_p\leftarrow$ ESPRIT algorithm for $\mathbf{\tilde{x}}_{b}^p, p=1,2,\cdots,P$
\State \quad $\hat{\gamma} \leftarrow$ $\frac{1}{P}\sum_{p=1}^{P}\hat{\gamma}_p$
\State \quad $\mathbf{R}_{\mathbf{\tilde{x}}_{c}}\leftarrow\mathbb{E}\left\{\mathbf{\tilde{x}}_{c}{\mathbf{\tilde{x}}_{c}^\mathrm{H}}\right\}$
\State \quad $\textbf{Q}$, $\bm{\Lambda}_\leftarrow$ decompose $\mathbf{R}_{\mathbf{\tilde{x}}_{c}}$ such that $\mathbf{Q}\mathbf{\Lambda}
\mathbf{Q}^\mathrm{H}$$=$$\mathbf{R}_{\mathbf{\tilde{x}}_{c}}$
\State \quad $\lambda_{max}\leftarrow\max\{\lambda_p|p=1,2,\cdots,P\}$
\State \quad $\mathbf{q}\leftarrow$the column of $\textbf{Q}$ corresponding to $\lambda_{max}$
\State \quad $\mathbf{q}_1$, $\mathbf{q}_2\leftarrow$ the first and the last $P-1$ elements of $\mathbf{q}$
\State \quad $e^{i\hat{\xi}}\leftarrow{\mathbf{q}_{1}}^\dagger\mathbf{q}_{2}$
\State \quad $\hat{\varphi}\leftarrow\arccos\sqrt{\big(\frac{\hat{\xi}-\hat{r}}{R}\big)^2+\cos{\hat{\gamma}}^2}$
\State \quad $\hat{\theta}\leftarrow\arctan\frac{\hat{\xi}-\hat{r}}{R\cos{\hat{\gamma}}}$
\State \quad \Return $\varphi$ and $\theta$
\State \textbf{end procedure}
\end{algorithmic}
\end{algorithm}

\subsection{Estimation of $\varphi$ and $\theta$}

Having estimated $r$ and $\gamma$, we can obtain the azimuth angle $\varphi$ and the elevation angle $\theta$. According to \eqref{xlk}, the signal received by the reference element $(\varphi_1=0)$ of the receive UCA on the zero OAM mode and $p$th frequency can be simplified as
\begin{align}
\tilde{x}'(k_p)&=-\frac{x_1(0,k_p)}{|x_1(0,k_p)|}\frac{s^*(0,k_p)}{|s(0,k_p)|}{\rm sign}(x_1(0,k_p))\nonumber\\
&=e^{ik_p\xi}+\tilde{z}'(k_p),
\label{amplitude}
\end{align}
where $\xi=r+R\sin\theta\cos\varphi$ is an intermediate parameter related to $\varphi$ and $\theta$, $\tilde{z}'(k_p)$ is the corresponding noise.

Then, the signals received on the $P$ frequencies can be collected in the vector
\begin{equation}\label{xn0}
\mathbf{\tilde{x}}_{c}=[\tilde{x}'(k_1), \tilde{x}'(k_2), \cdots, \tilde{x}'(k_P)]^\mathrm{T},
\end{equation}
which can be expressed in a compact form as
\begin{equation}
\mathbf{\tilde{x}}_{c}=\mathbf{c}+\mathbf{z}_{c},
\end{equation}
where $\mathbf{c}$ $=$ $[e^{ik_1\xi}, e^{ik_2\xi},\cdots, e^{ik_P\xi}]^T$ and $\mathbf{z}_{c}$ is the corresponding noise vector.

The covariance matrix of $\mathbf{\tilde{x}}_{c}$ can be written as
\begin{equation}
\mathbf{R}_{\mathbf{\tilde{x}}_{c}}=\mathbb{E}\left\{\mathbf{\tilde{x}}_{c}{\mathbf{\tilde{x}}_{c}}^H\right\}
=\mathbf{c}\mathbf{c}^H+\mathbf{R}_{\mathbf{z}_{c}},
\end{equation}
where $\mathbf{R}_{\mathbf{z}_{c}}=\mathbb{E}\left\{\mathbf{z}_{c}{\mathbf{z}_{c}}^\mathrm{H}\right\}$. The eigenvalue decomposition (EVD) of $\mathbf{R}_{\mathbf{\tilde{x}}_{c}}$ is
\begin{equation}\label{evd}
\mathbf{R}_{\mathbf{\tilde{x}}_{c}}=\mathbf{Q}\mathbf{\Lambda}\mathbf{Q}^{H},
\end{equation}
where $\mathbf{Q}$ is an $P\times P$ unitary matrix and $\mathbf{\Lambda}=\textrm{diag}\{\lambda_1,$ $\lambda_2,\cdots,\lambda_{P}\}$. Denote as $\lambda_{max}=\max\{\lambda_p|p=1,2,\cdots,P\}$, and as $\mathbf{q}$ the eigenvector corresponding to $\lambda_{max}$, so that it is
\begin{align}
\mathbf{R}_{\mathbf{\tilde{x}}_{c}}\mathbf{q}=\lambda_{max}\mathbf{q}.
\end{align}
Now, the subspace spanned by $\mathbf{q}$ is the signal subspace spanned by $\mathbf{c}$ so that the following relationship holds true
\begin{equation}
\mathbf{c}=\textrm{T}\mathbf{q},
\end{equation}
where $\textrm{T}$ is a non-zero parameter. If we consider the two vectors $\mathbf{c}_{1}$ and $\mathbf{c}_{2}$, obtained by taking the first and the last $P-1$ elements of $\mathbf{c}$, respectively, it is $\mathbf{c}_{2} = {\Phi}\mathbf{c}_{1}$, where $\Phi=e^{i\xi}$. To obtain the estimate of $\xi$, we construct the two vectors $\mathbf{q}_{1}$ and $\mathbf{q}_{2}$, composed by the first  and  by the last $P-1$ elements of $\mathbf{q}$, respectively. Then, exploiting the fact that $\mathbf{c}_{1}=\textrm{T}\mathbf{q}_{1}$ and $\mathbf{c}_{2}=\textrm{T}\mathbf{q}_{2}$, one obtains
\begin{equation}
\mathbf{q}_{2} = {\Phi}\mathbf{q}_{1},
\end{equation}
which leads to
\begin{equation}
e^{i\hat{\xi}}={\mathbf{q}_{1}}^\dagger\mathbf{q}_{2},
\end{equation}
where $\hat{\xi}$ is the estimate of $\xi$ based on the signal vector $\mathbf{\tilde{x}}_{c}$.

After that, according to
\begin{align}
&\hat{\varphi}=\arccos\sqrt{\big(\frac{\hat{\xi}-\hat{r}}{R}\big)^2+\cos{\hat{\gamma}}^2},\nonumber\\
&\hat{\theta}=\arctan\frac{\hat{\xi}-\hat{r}}{R\cos{\hat{\gamma}}},
\end{align}
the estimates of the azimuthal angle $\varphi$ and the elevation angle $\theta$ can be obtained. Thus, the AoA estimation by M-F MT-ESPRIT algorithm is completed, and the detailed procedure is summarized in Algorithm 1. Compared with the existing AoA estimation method \cite{Chen2020Multi}, the AoA estimation method based on M-F MT-ESPRIT algorithm can estimate the azimuth and elevation angles of muiti-mode OAM beams only by utilizing the exponentials of the received training signals, which is easier to implement in practical applications.

\begin{table}[tb]
\caption{The complexity of the AoA estimation.}
\setlength{\tabcolsep}{0.5mm}{
\begin{tabular}{ll}
  \toprule
  \qquad\qquad\textbf{Procedure}                                                                                  &\textbf{\qquad\qquad Complexity}\\
  \midrule
  \quad Estimating $\hat{r}$                                                                                      &\qquad\quad$\mathcal{O}\left(UP^{\max\{2,g\}}\right)$\\
  \quad Estimating $\hat{\gamma}$                                                                                 &$\qquad\quad\mathcal{O}\left(PU^{\max\{2,g\}}\right)$\\
  \quad Eq.$(11):\mathbf{R}_{\mathbf{\tilde{x}}_{c}}=\mathbb{E}\left\{\mathbf{\tilde{x}}_{c}{\mathbf{\tilde{x}}_{c}}^H\right\}$           &$\qquad\qquad\quad\mathcal{O}\left(P^2\right)$\\
  \quad Eq.$(12):\mathbf{R}_{\mathbf{\tilde{x}}_{c}}=\mathbf{Q}\mathbf{\Lambda}\mathbf{Q}^{H}$                            &$\qquad\qquad\quad\mathcal{O}\left(P^g\right)$\\
  \quad Eq.$(16):e^{i\hat{r}_u}={\mathbf{q}^{1}_u}^\dagger\mathbf{q}^{2}_u$                                       &$\qquad\qquad\quad\mathcal{O}\left(P\right)$\\
  \midrule
  \multicolumn{2}{l}{The total complexity: $\mathcal{O}\left(UP^{\max\{2,g\}}\right)+\mathcal{O}\left(PU^{\max\{2,g\}}\right)$}  \\
  \bottomrule
  \label{Table1}
\end{tabular}}
\vspace{-3.5mm}
\end{table}

\begin{figure}[t] %figure1
\setlength{\belowcaptionskip}{0.3cm}   %调整图片标题与下文距离
\begin{center}
\includegraphics[width=9.0cm,height=6.7cm]{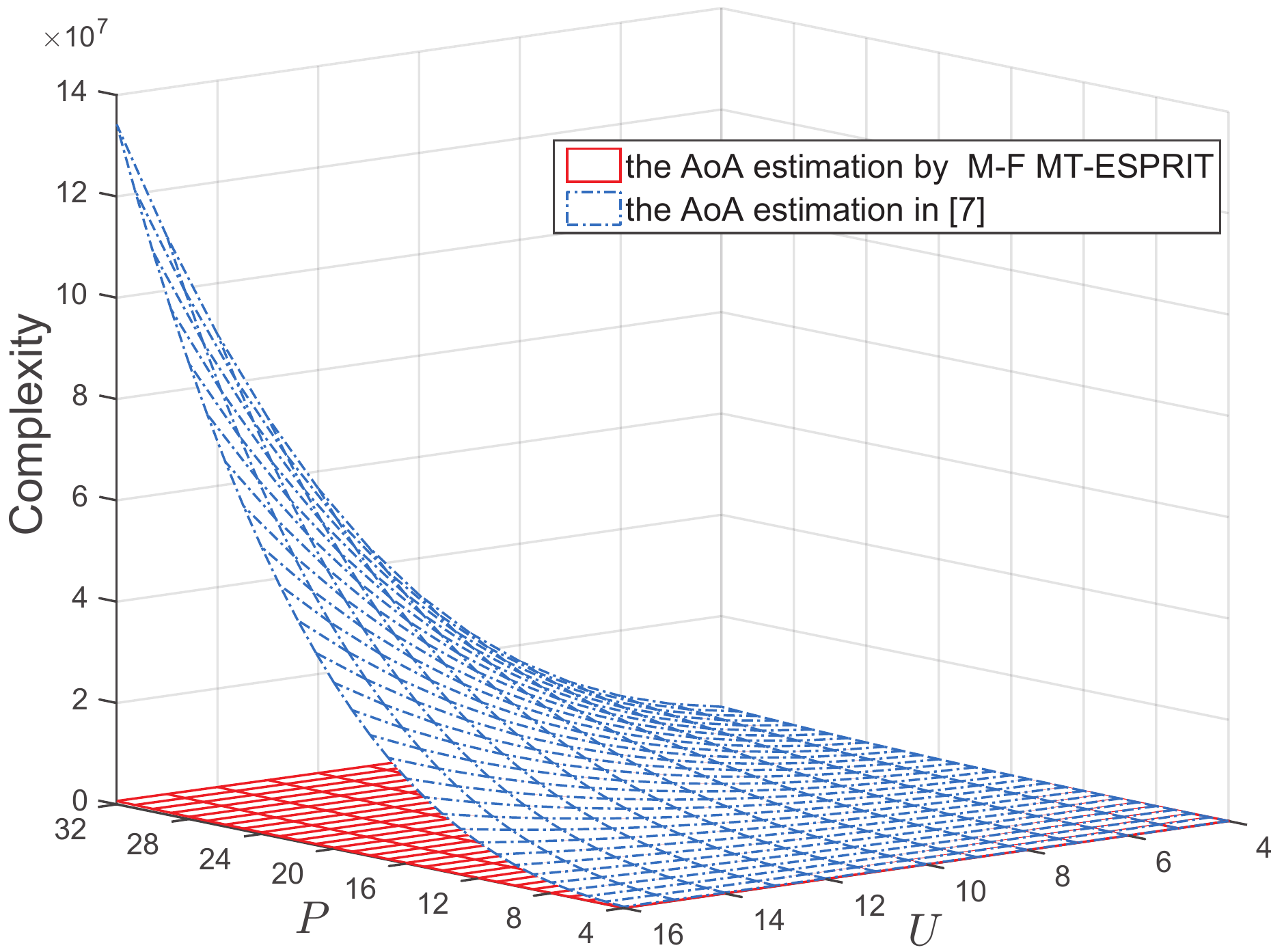}
\end{center}
\caption{The complexities of the AoA estimation method based on M-F MT-ESPRIT algorithm and the AoA estimation method in \cite{Chen2020Multi} vs. $P$ and $U$ with $g=3$.}
\label{Fig8}
\end{figure}

The computational complexities of the proposed AoA estimation method based on M-F MT-ESPRIT algorithm is summarized in Table \ref{Table1}, where $g$ is generally equal to 3, but can be reduced to 2.376 when the Coppersmith and Winograd algorithm \cite{pan1999complexity} is applied to the eigenvalue decomposition in \eqref{evd}. Due to the use of M-F MT-ESPRIT algorithm, the dimension of the training signal vector processed by ESPRIT algorithm each time is greatly reduced. The total complexity comparison between the existing AoA estimation method \cite{Chen2020Multi} and the proposed AoA estimation method based on M-F MT-ESPRIT algorithm is shown in Fig.\ref{Fig8}.

\begin{figure}[t]
\setlength{\abovecaptionskip}{0cm}   %调整图片标题与图距离
\setlength{\belowcaptionskip}{-0.2cm}   %调整图片标题与下文距离
\begin{center}
\includegraphics[width=8.3cm,height=7.5cm]{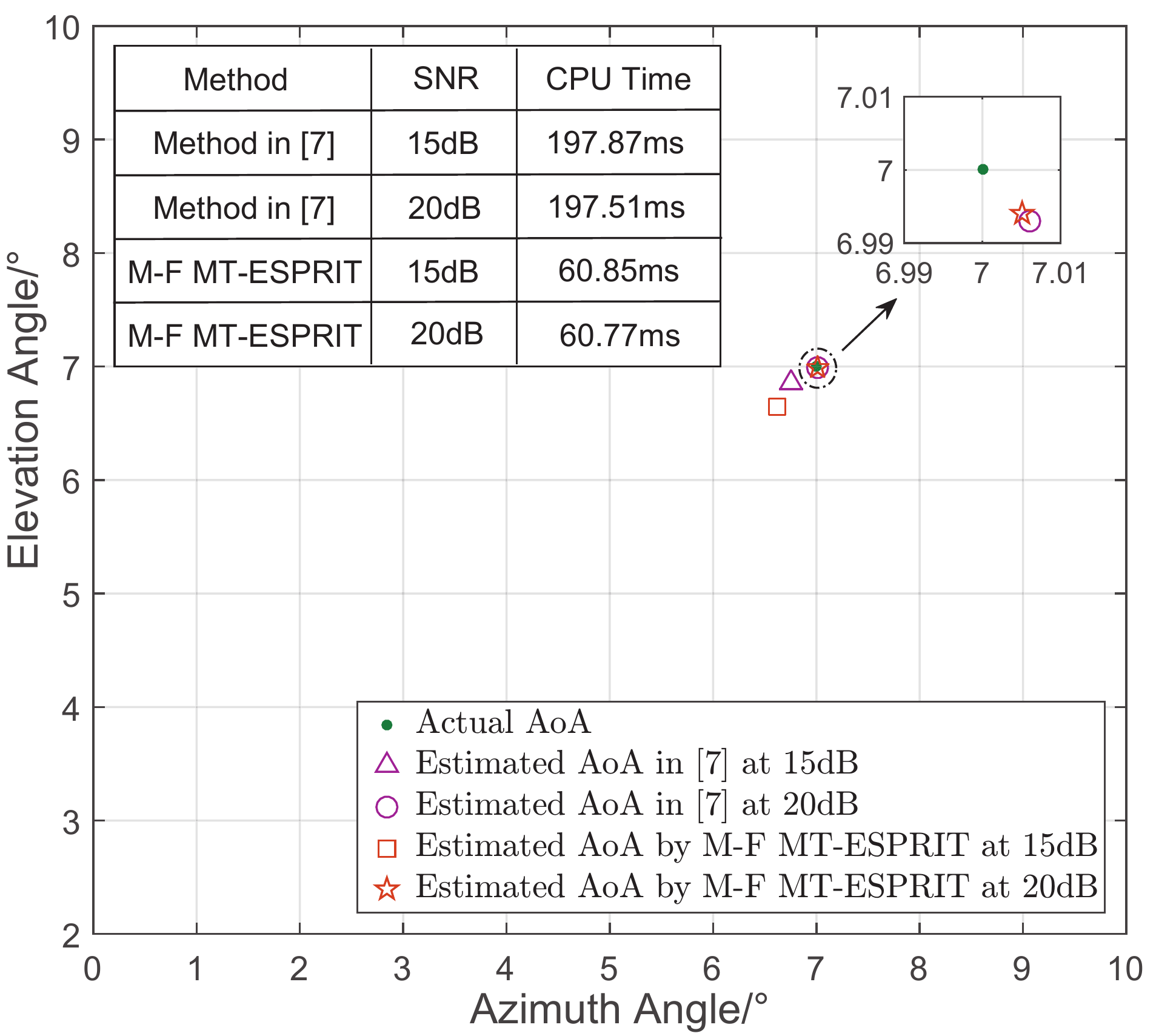}% scale=0.42
\end{center}
\caption{The AoA estimation results at 15dB and 20dB.}
\label{Fig5}
\end{figure}

\vspace{1.5mm}
\section{Simulation Results and Performance Analysis}
\vspace{2.5mm}

We first verify the proposed AoA estimation method by numerical simulations, and then compare the accuracy of the AoA estimation method proposed in this paper with that proposed in \cite{Chen2020Multi}. Thereafter, we compare the capacity performance of the OAM channel without beam steering and that with beam steering \cite{Chen2020Multi} based on the estimated AoA. In the simulation, we choose $N=9$, $P = 8$ subcarriers from 2.244GHz to 2.578GHz corresponding to the wave number $k_p = 47, 48,\dots, 54$, number of OAM modes $U= 8$ with $\ell_u = -4,-3,\cdots,+3$, $R=10\lambda_1$, $\lambda_1=2\pi/k_1$, $(r, \varphi, \theta) = (40\textrm{m}, 7^{\circ}, 7^{\circ})$. Besides, we assume the angle range of receive UCA's main lobe $[\alpha_a,\alpha_b]=[2^{\circ},8^{\circ}]$, the initial number of intervals $\mathcal{D}=2$ \cite{Chen2020Multi}. Unless otherwise stated, the SNRs in all the figures are defined as the ratio of the received signal power versus the noise power.

The AoA estimation results of the proposed M-F MT-ESPRIT method and the method proposed in \cite{Chen2020Multi} are shown in Fig. \ref{Fig5}. It can be seen from the figure that with the increase of SNR the estimated AoAs approach to the actual value. When SNR$=20$dB, the AoA estimated by M-F MT-ESPRIT method is $(\hat{\varphi}, \hat{\theta}) = (7.005^{\circ}, 6.994^{\circ})$, and the AoA estimated by the method in \cite{Chen2020Multi} is $(\hat{\varphi}, \hat{\theta}) = (7.006^{\circ}, 6.993^{\circ})$, which are very close to the actual AoA. Besides, the CPU time of M-F MT-ESPRIT method is much lower than that of the method in \cite{Chen2020Multi}.

\begin{figure}[t] %figure1
\begin{center}
\includegraphics[width=8.4cm,height=6.9cm]{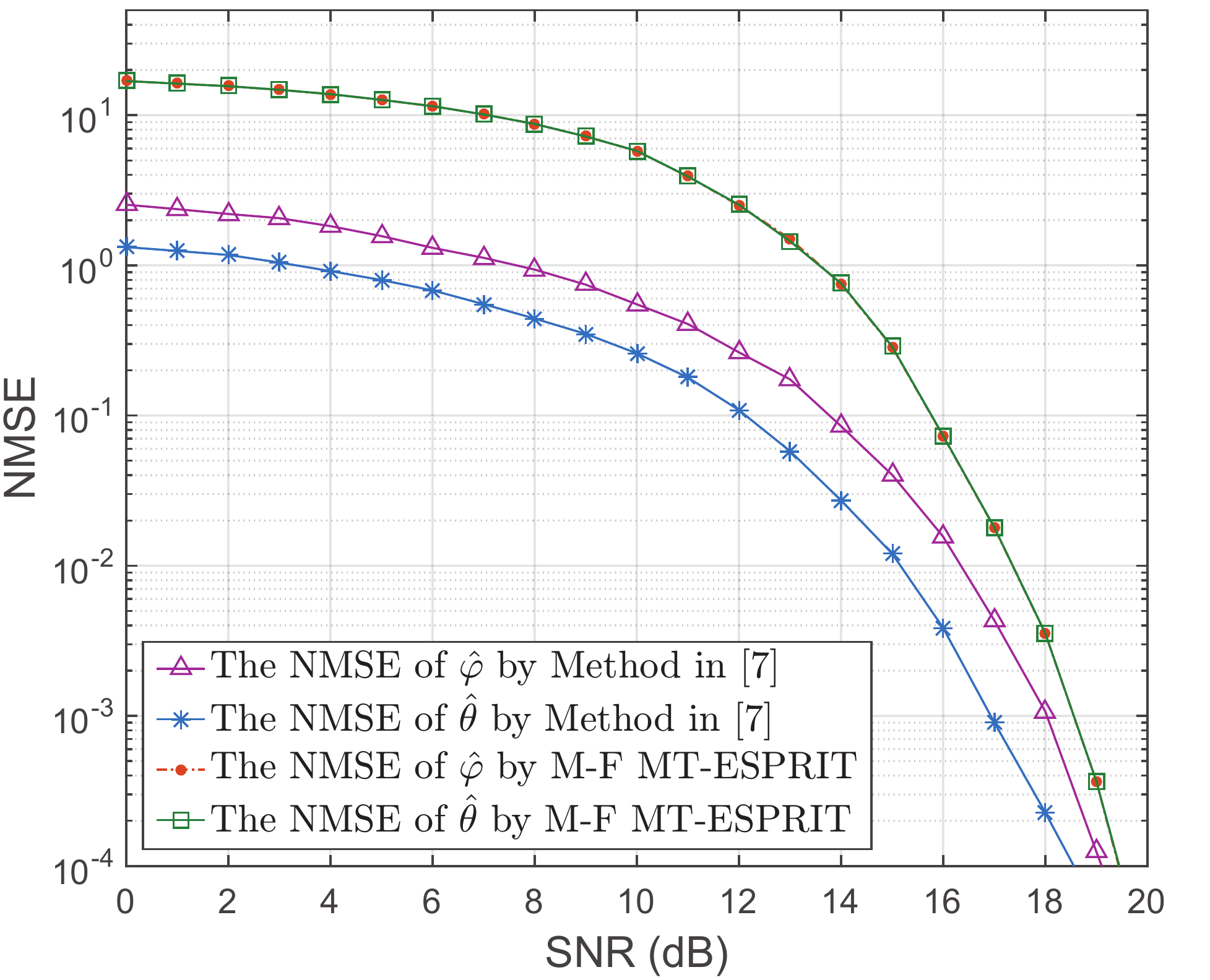}
\end{center}
\caption{The NMSEs of the estimated $\hat{\varphi}$ and $\hat{\theta}$ vs. SNR.}
\label{Fig6}
\end{figure}
\begin{figure}[t] %figure1
\begin{center}
\includegraphics[width=8.0cm,height=7.0cm]{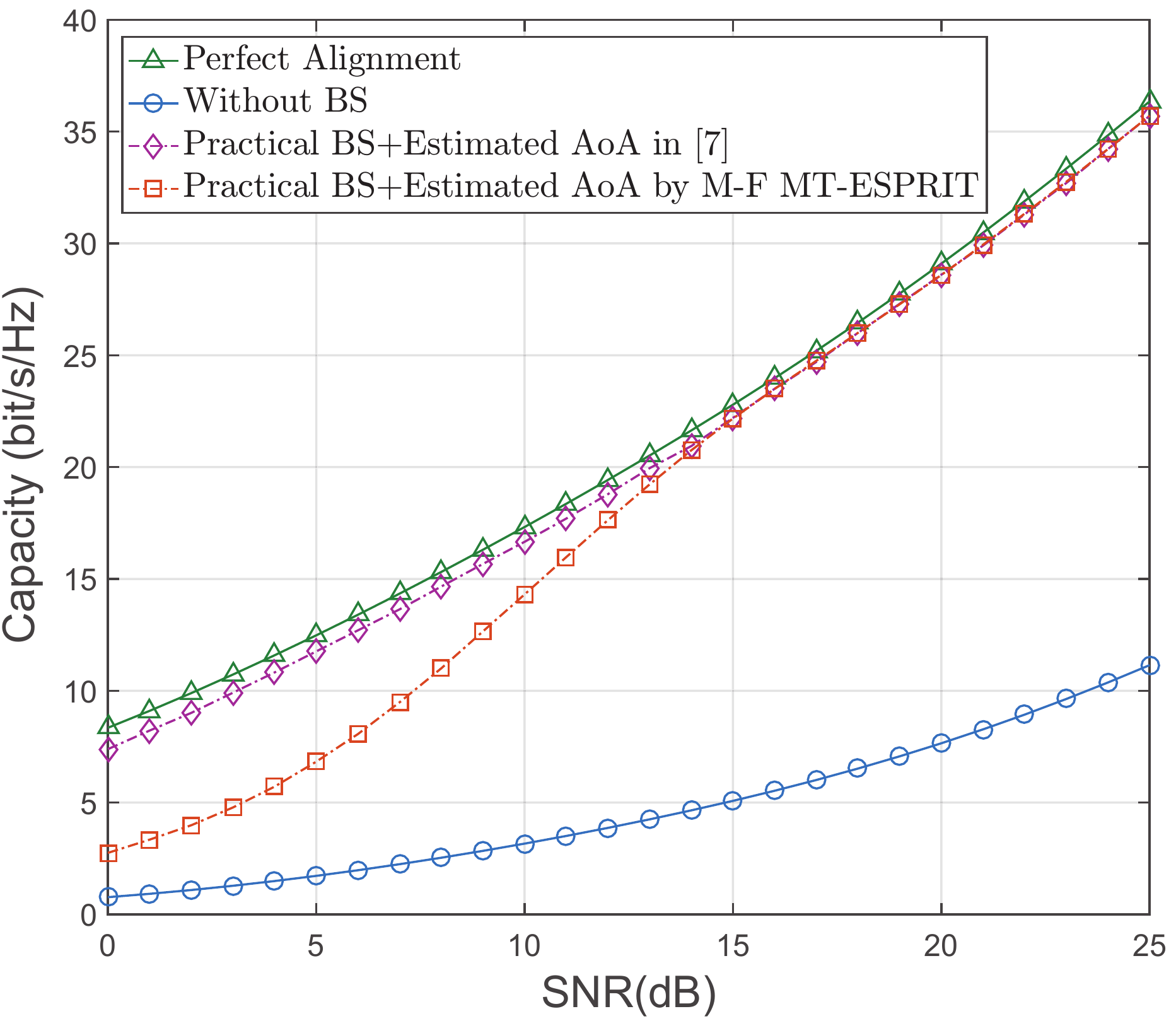}
\end{center}
\caption{The capacity comparison of the OAM channel with perfect alignment, without BS, with BS by the estimated AoA in this paper, and with BS by the estimated AoA in \cite{Chen2020Multi}. BS: beam steering.}
\label{Fig7}
\end{figure}
Fig.\ref{Fig6} illustrates the normalized mean-squared errors (NMSEs) of $\hat{\varphi}$ and $\hat{\theta}$ obtained by M-F MT-ESPRIT method and the method in \cite{Chen2020Multi}. The NMSE is defined as $\mathbb{E}\{(\hat{x}-x)^2/x^2\}$, where $\hat{x}$ denotes the estimate of $x$. For the proposed M-F MT-ESPRIT method, as SNR increases the NMSEs of $\hat{\varphi}$ and $\hat{\theta}$ decline. It follows that the proposed M-F MT-ESPRIT method is able to accurately estimate the AoA of multi-mode OAM beams. Moreover, due to the reduction of the dimension of the training signal vector processed by ESPRIT algorithm each time, the NMSEs of $\hat{\varphi}$ and $\hat{\theta}$ obtained by M-F MT-ESPRIT method is higher than those obtained by the method in \cite{Chen2020Multi}.

In Fig.\ref{Fig7}, we compare the capacity of the OAM channel with perfect alignment, without beam steering, with beam steering by the estimated AoA in this paper, and with beam steering by the estimated AoA in \cite{Chen2020Multi}. The capacity of the misaligned OAM channel is greatly improved and approaches the capacity of accurately aligned OAM channel after applying the beam steering with the AoA estimation method proposed in this paper and in \cite{Chen2020Multi}. Therefore, the accurate AoA estimation is essential for the OAM receiver to improve the capacity of the misaligned OAM channel. Moreover, due to the larger estimation error of $\hat{\varphi}$ and $\hat{\theta}$, the capacity of the OAM channel with beam steering by the estimated AoA in this paper is lower than that by the estimated AoA in \cite{Chen2020Multi} under the low SNRs.
%beam steering based on the estimated AoA can solve the problem of misalignment between the transmit and the receive UCAs.

\section{Conclusions}
In this paper, an AoA estimation method based on M-F MT-ESPRIT algorithm for OAM wireless communication systems is proposed, which can accurately estimate the azimuth and elevation angles of multi-mode OAM beams only based on the received training signals. With the accurately estimated AoA, the capacity of the misaligned OAM channel can be greatly improved by beam steering and approaches the capacity of accurately aligned OAM channel. Moreover, compared with the existing AoA estimation method \cite{Chen2020Multi}, the total complexity of the proposed AoA estimation method based on M-F MT-ESPRIT algorithm is significantly decreased, which is more practical.

\section*{Acknowledgment}
The authors would like to thank the editor and the anonymous reviewers for their careful reading and valuable suggestions that helped to improve the quality of this manuscript.

\bibliographystyle{IEEEtran}
\bibliography{IEEEabrv,single_user_aoa}
 \end{document}